\documentclass[prl,twocolumn,superscriptaddress,showpacs]{revtex4}
\usepackage{epsfig,times,amsmath,amssymb,color}
\begin{document}

\title{Exotic paired states with anisotropic spin-dependent Fermi surfaces}
\author{Adrian E. Feiguin}
\affiliation{Microsoft Project Q, University of California,  Santa Barbara,  CA 93106, USA}
\affiliation{Condensed Matter
Theory Center, Department of Physics, University of Maryland,
College Park, Maryland 20742-4111, USA.}
\author{Matthew P. A. Fisher} 
\affiliation{Microsoft Project Q, University of California,  Santa Barbara,  CA 93106, USA}

\date{\today}

\begin{abstract}
We propose a model for realizing exotic paired states in cold atomic Fermi gases. By using a {\it spin dependent} optical lattice it is possible to engineer spatially anisotropic Fermi surfaces for each hyperfine species, that are rotated 90 degrees with respect to one another. We consider
a balanced population of the fermions with an attractive interaction.
We explore the BCS mean field phase diagram as a function of the anisotropy, density, and interaction strength,
and find the existence of an unusual paired superfluid state with coexisting pockets of momentum space with gapless unpaired carriers. This state is a relative of the Sarma or breached pair states in polarized mixtures, but in our case the Fermi gas is unpolarized. 
We also propose the possible existence of an exotic paired ``Cooper-pair Bose-Metal" (CPBM) phase, which 
has a gap for single fermion excitations but gapless and uncondensed ``Cooper pair" 
excitations residing on a ``Bose-surface" in momentum space.
 \end{abstract}
\pacs{74.20.-z, 74.25.Dw, 03.75.Lm} 
\maketitle


Fermionic superfluidity is, with its many manifestations ranging from 
solid state physics
to ultracold atom gases and neutron stars, one of the most fascinating 
phenomena. 
In recent years we have experienced a renewed interest in some of the basic questions behind this problem, motivated by technical advances in the control and manipulation of cold atomic gases.\cite{Z-K-review}
In particular, a great deal of attention is being devoted to the quest for unconventional paired states of matter.

Conventional superconductivity (BCS theory \cite{BCS}) consists of 
pairing between fermions with opposite spin and equal but opposite 
momentum.    In the presence of (partial) spin polarization more unusual superfluid states are possible.
In the 
Fulde-Ferrell-Larkin-Ovchinnikov (FFLO) state\cite{FF,LO}
pairing occurs across the mismatched 
Fermi surfaces so that the Cooper pairs have a finite center-of-mass 
momentum. \cite{FFLOreviews}  The excess unpaired fermions occupy
a Fermi sea, so that there are gapless fermionic excitations coexisting with superfluidity.
The FFLO state has been observed in solids only recently \cite{Radovan Nature 03}. 
An interesting alternative to the FFLO state in partially polarized systems is astate with zero momentum pairing as suggested by Sarma \cite{sarma,breached1,breached2,rupak,caldas}.   
With disconnected regions of momentum space with pairing, separated by an unpaired polarized sea of fermions, the 
Sarma state is sometimes referred to as a 
 ``breached  pair" phase \cite{breached1,breached2}).
Recently a ``pair density wave" or striped superconducting state 
with a spatially modulated order parameter similar to the FFLO state but without time-reversal
symmetry breaking has been proposed to account for experimental observations in La$_{2-x}$Ba$_x$CuO$_4$. \cite{Berg08,Berg09} 

These unconventional paired states might be accessible in cold atom experiments, with ``polarized" Fermi gases in two hyperfine states having different populations. \cite{Z-K-review,experiments1,experiments2}  
Additionally, one can 
explore the effects of differing masses in Fermi mixtures with two atomic species.\cite{mass imbalance}
A mass ``imbalance" between the two hyperfine states of a single atomic species
can also be achieved in an optical lattice by tuning a {\it spin-dependent} hopping with light.
Indeed, spin-dependent optical lattices have already been demonstrated \cite{spin-dependent}.
A  protocol to realize a Hubbard model with spin-dependent hopping has been proposed by Liu {\it et al.} in Ref.\cite{liu04}, by tuning the lattice lasers between fine structure levels of $^{40}$K atoms, and using Feschbach resonances to manipulate the interactions.
By varying the polarization and mass ratios 
it might be possible to realize unconventional paired states such as the FFLO or Sarma state.

\begin{centering}
\begin{figure}
\epsfig {file=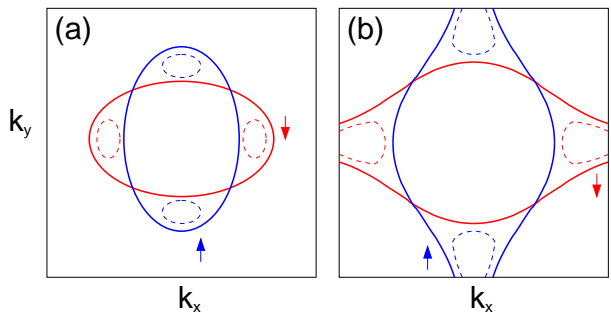,width=80mm} 
\caption{
(color online) Schematic Fermi surfaces for the dispersions $\epsilon_\sigma(k_x,k_y)$ described by 
Eq.(\ref{omega}): (a) closed Fermi surfaces for small density; (b) open 
Fermi surfaces for higher density/large anisotropy. Arrows label two hyperfine states of the same atomic species. Dashed lines delimit the pockets of the momentum distribution in the gapless G-SF solution. 
} \label{fermi surface}
\end{figure}
\end{centering}

In this paper we suggest a different means to access unconventional paired states
in ultracold atomic systems. 
Our idea is to create mismatched Fermi surfaces in an {\it unpolarized} mixture.
Consider an experiment with two hyperfine states (that we label $\uparrow$ and $\downarrow$ hereafter) with equal population that are moving, for simplicity, on a two-dimensional square lattice.  Imagine tuning the hopping so that one spin state hops preferentially along the $x$-axis and the other preferentially along the $y$-axis.  
We are interested in the situation where the Fermi surfaces of the two spin states
are rotated by $90$ degrees with respect to one another, and focus on a near neighbor hopping  Hamiltonian with single particle dispersions,
\begin{eqnarray}
\epsilon_\uparrow(k_x,k_y)=-2t_a\cos{(k_x)}-2t_b\cos{(k_y)} - \mu,\\
\epsilon_\downarrow(k_x,k_y)=-2t_b\cos{(k_x)}-2t_a\cos{(k_y)} - \mu,
\label{omega}
\end{eqnarray}
with chemical potential $\mu$.
The ratio $t_b/t_a \le 1$ determines the 
eccentricity of the two Fermi surfaces (see Fig.\ref{fermi surface}). 
A similar Fermi surface geometry has been proposed for the ruthenate materials, \cite{raghu0902.1336,lee0902.1337} arising in a multi-band model which possesses spin nematic order \cite{oganesyan01,Wu07}.   The nematic behavior originates from hybridization of the bands and is a spontaneous symmetry breaking, while our Hamiltonian explicitly breaks time-reversal and rotational symmetry.  

As we discuss below, in the presence of an attractive interaction this peculiar arrangement of anisotropic Fermi surfaces predicts a new class of 
gapless superfluid (G-SF) states within a BCS mean field theory, and might 
also give rise to an even more exotic paired but {\it non-superfluid} Bose-Metal phase. 

To be concrete we assume the particles interact through a short-range s-wave 
potential, that we represent using the attractive Hubbard model: 
\begin{eqnarray}
H =  & \sum\limits_{k,{\sigma}} \epsilon_\sigma(k) c^\dagger_{k,\sigma}
c_{k,\sigma} + 
U \sum\limits_{i} n_{i,\uparrow}n_{i,\downarrow} 
~, \label{one}
\end{eqnarray}
where $c^\dagger_{k,\sigma}$ creates a fermion with spin
$\sigma=\,\uparrow ,\downarrow$ at momentum $k$,
$n_{i\sigma}=c^\dagger_{i\sigma}c_{i \sigma}$ is the local 
on-site density, and $U$ is the
interaction strength that we will take to be negative
(attractive). 

By simple inspection of the spin-dependent Fermi surfaces in 
Fig.\ref{fermi surface}, we can anticipate several different pairing 
possibilities within a BCS treatment.
For a very strong attractive interaction, $|U|>> t_a,t_b$, a state with zero  
momentum pairing and a fully gapped Fermi surface is expected.    
For smaller $U$ a zero-momentum paired state with 
gapless single fermion excitations analogous
to the Sarma state in polarized mixtures could occur. 
Alternatively, pairing could occur across the 
two mismatched Fermi surfaces with a finite momentum condensate,
an unpolarized analog of the FFLO state.

We now implement a self-consistent BCS approximation
to obtain the mean-field phase diagram and explore the properties
of the resulting paired superfluid states.  A discussion of more exotic non-superfluid paired phases
which are inaccessible 
within mean field theory will be presented below.
We consider an on-site pairing of the form $-U/V\langle c_{j\downarrow}c_{j\uparrow}\rangle = 
\Delta \exp{(iQj)}$  with gap parameter $\Delta$.  Here $V=L^2$ is the number of sites in the system and
$j$ denotes the center of mass position of a pair in 
real space. The resulting quadratic Hamiltonian can be readily 
diagonalized, yielding quasi-particle excitations with dispersion:
\begin{equation}
E_{\pm}(k) = \sqrt{\epsilon_+(k)^2 + \Delta^2} \pm \epsilon_-(k),
\label{dispersion}
\end{equation}
where we have defined,
\begin{equation} \epsilon_{\pm}(k) =  [\epsilon_\uparrow(-k+\frac{Q}{2}) \pm \epsilon_\downarrow(k+\frac{Q}{2})]/2 .
\end{equation}


Consider first the case of zero pairing momentum $Q=0$. Besides the 
normal phase with $\Delta=0$, we expect a fully gapped BCS-like solution when $E_\pm(k) >0$
for all momenta in the Brillouin zone (BCS state).  This occurs 
when the attractive interaction is sufficiently strong.   For smaller $\Delta$
a gapless superfluid phase (G-SF)  is also possible.\cite{breached1,caldas}
Gapless fermion excitations occur at momenta that satisfy
$E_+(k)=0$ or $E_-(k)=0$, or equivalently, 
\begin{equation}
\epsilon_\uparrow(k)\epsilon_\downarrow(k) = -\Delta^2.
\label{deltac}
\end{equation}
The resulting closed curves in momentum space are sketched schematically as
dashed lines in Fig. 1.  In the case of closed Fermi surfaces as shown in Fig. 1(a) there are four small pockets near the Fermi points $(\pm k_{Fx}^\downarrow,0)$ and $(0,\pm k_{Fy}^\uparrow)$.
The solution with open Fermi surfaces (see Fig.1(b)) has two large pockets that wrap around the Brillouin zone.  

It is instructive to compute the momentum distribution function for
the two spin species,  $n_\sigma(k) \equiv \langle c^\dagger_{k,\sigma} c_{k,\sigma}\rangle$.
There are three distinct regions in momentum space determined by the conditions: (i)  $E_-(k)>0$ and $E_+(k) < 0$
which occurs inside the $\uparrow$ pockets; (ii) $E_+(k) > 0$ and $E_-(k) < 0$ which occurs inside the $\downarrow$ pockets; or (iii) both $E_\pm(k) > 0$  which occurs over the rest of the Brillouin zone.
For momentum where (i) is satisfied we obtain $n_\uparrow(k) = 1, n_\downarrow(k) = 0$,
whereas in region (ii) one has $n_\uparrow(k) = 0, n_\downarrow(k) = 1$.
Thus, inside these pockets the fermions are completely unpaired,
and there is no mixing between the two spin species.  Finally, in region (iii) the fermions are
completely paired and one obtains; 
\begin{equation}
n_\uparrow(k)=n_\downarrow(k) = \frac{1}{2}\left(1-\frac{\epsilon_+(k)}{\sqrt{\epsilon_+(k)^2+\Delta^2}}\right).
\end{equation}
The equivalence of the two momentum distribution functions and the restoration
of the four fold symmetry of the square lattice are presumably artifacts of BCS  mean-field theory,
and would be modified when fluctuations are taken into account.
Even within BCS mean-field theory, the spectral functions of the two fermion species are different from one another.    

While possessing 
regions in momentum space with complete pairing and other regions with no pairing whatsoever
is reminiscent of the Sarma phase, this novel {\it unpolarized} gapless superfluid (G-SF) phase 
has unpaired fermion pockets of {\it both} spin species.


We now analyze the (mean-field) stability of the normal, BCS and gapless superfluid   (G-SF) phases, by solving the BCS saddle point (gap) equations numerically on 
the square lattice.  We work in the canonical ensemble, with fixed density, which is the relevant situation for cold fermionic gases, also allowing for finite momentum pairing with momenta $Q$ either along the $x,y$-axes or along
the diagonals in the Brillouin zone.
The Helmholtz free energy in the canonical ensemble is computed and minimized in systems sizes up to $300 \times 300$ sites following the prescription detailed in Ref.[\onlinecite{koponen06}].
 
\begin{centering}
\begin{figure}
\epsfig {file=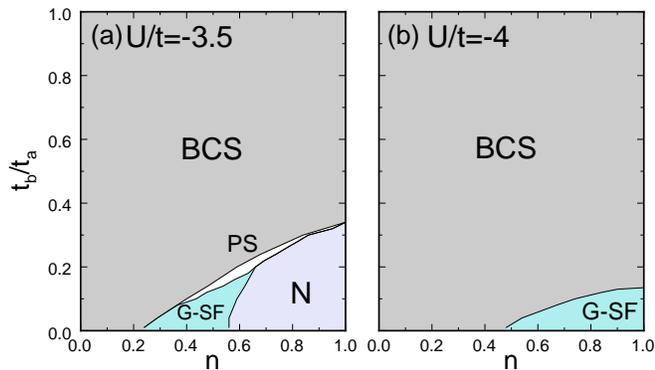,width=50mm,angle=-90} 
\caption{
Mean-field phase diagram of the attractive Hubbard model as a function of the anisotropy $t_b/t_a$ and fermion density $n$, for values of the interaction (a) $U=-3.5t$ and (b) $U=-4t$. We find regions occupied by a fully paired solution with $Q=0$ (BCS), normal (N), gapless superfluid (G-SF), and phase separation (PS).
}
\label{phase diagram}
\end{figure}
\end{centering}

 For small values of the attraction $|U|/t$ we find that the system is in the normal state, independently of the total fermion density, $n$, or the value of the anisotropy $t_b/t_a$, while for large values of the interaction strength, the system is in a fully gapped superfluid state with $Q=0$ (BCS).
 The interesting regime is for values of $2t \lesssim |U| \lesssim 4t$, where we find a competition of phases.
 In Fig.\ref{phase diagram} we show the phase diagram for two different values of $|U|/t$, as a function of the fermion density and the anisotropy $t_b/t_a$. 
 At $|U|/t=3.5$, the normal phase occupies a region at high densities and high anisotropy, while the superconducting phase (BCS) dominates at intermediate and small anisotropy. Separating these two phases is a window where the G-SF state is realized. There is also a tiny sliver where phase separation takes place, that coincides with the onset of the instability toward the G-SF state. We determined the boundaries of this region by identifying densities with negative compressibility $dn/d\mu$, and using the Maxwell construction. Contrary to the Sarma phase in polarized mixtures, the G-SF state is stable in a wide region of parameters.\cite{rupak,caldas}
 At larger values of the interaction the normal phase becomes energetically unfavorable,
 leaving a region of the G-SF phase that shrinks with increasing interaction. 
In all cases the G-SF phase only occurred in the with open Fermi surfaces/large pockets (see Fig.1(b)).
We also looked for energy minima at finite momentum,
but we have not found solutions for the system sizes and in the parameter space that we studied.

We now turn to the possibility of paired states which are {\it not superfluids}.
This is motivated by recent work\cite{DBL}
which introduced and studied a hard core boson model with ring exchange, and
made strong arguments for the existence of a novel Bose-Metal phase,
a D-wave Bose Liquid.
Here we will refer to this phase as a D-wave Bose-Metal (DBM).
The proposed Bose-Metal phase is not a superfluid,
having zero condensate and superfluid densities, but possesses 
gapless excitations which reside on surfaces
in momentum space.  In contrast to a conventional Fermi liquid, in the DBM phase
the gapless excitations cannot be described in terms of weakly interacting
quasiparticles.  The low energy excitations are intrinsically strongly interacting fluctuations, and a single particle description is incorrect.
In the DBM phase pairs of {\it bosons} possess D-wave correlations,
but the phase is {\it not} a paired-boson condensate.

\begin{centering}
\begin{figure}
\epsfig {file=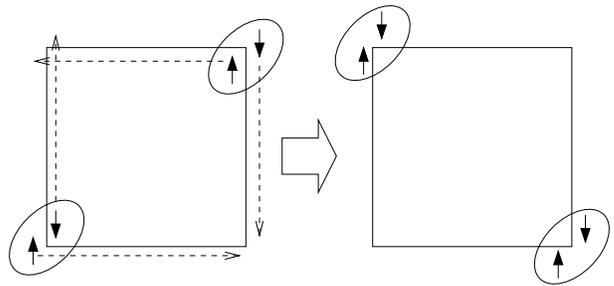,width=80mm} 
\caption{
Processes that would give origin to an effective ring-exchange of pairs.
fermions of different hyperfine species (with opposite spin) prefer to move along perpendicular directions (left). 
}
\label{ring}
\end{figure}
\end{centering}

A connection between our model of spin-full fermions
with spin-dependent spatially anisotropic Fermi surfaces and the Bose-Metal phase explored in Ref.[\onlinecite{DBL}]  can be  made as follows.  First we define an on-site ``Cooper pair" operator,
$b^\dagger_i = c^\dagger_{i \uparrow} c^\dagger_{i \downarrow}$.
This operator is a ``hard-core" boson, commuting at different sites,
$[ b_i, b^\dagger_j]=0$ for $i \ne j$, yet satisfying $b^\dagger_i b^\dagger_i = 0$.
Now consider the regime with $|U| >> t_a,t_b$ where all of the fermions
are tightly bound into on-site Cooper pairs.  The goal is to derive an
{\it effective} Hamiltonian for these hard-core bosons, by considering
a perturbation expansion in powers of $t_a/|U|,t_b/|U|$.
At second order one obtains a boson hopping term $H_J$ with strength
$J \sim t_a t_b/|U|$, 
\begin{equation}
H_J = - J \sum_{\langle i,j\rangle} b^\dagger_i b_j + H.c.  .
\end{equation}
The hoping $J$ is proportional to a product of $t_a$ and $t_b$ since
it is necessary to hop both the $\uparrow$-spin and $\downarrow$-spin fermion in order to move the
hard-core boson.  
Notice that in the extreme anisotropic limit with $t_b=0$ and $t_a \ne 0$
it is actually impossible for a single boson to move, $J=0$.  

But consider the boson ring-exchange term as depicted in Fig.\ref{ring}, which involves two bosons on opposite corners
of an elementary square plaquette rotating by $\pm 90$ degrees,
\begin{equation}
H_{\mathrm{ring}} = K_{\mathrm{ring}} \sum_{\mathrm{plaquettes}} b^\dagger_1 b_2 b^\dagger_3 b_4 + H.c ,
\end{equation}
with $i=1,2,3,4$ labeling sites taken clockwise around the
square plaquette.    The ring-exchange coupling strength $K_{\mathrm{ring}}>0$ 
can be computed at fourth order in $t_{a,b}/|U|$.  
In the highly anisotropic limit with $t_a >> t_b$, the process depicted in Fig.\ref{ring} dominates,
giving $K_{\mathrm{ring}} \sim t_a^4/|U|^3$.  Remarkably, while the single boson hopping term vanishes in the extreme anisotropic limit, $t_b/t_a \rightarrow 0$,
the ring term which hops pairs of bosons is non-zero.
Thus, with increasing anisotropy, $t_a/t_b$, the ratio $K_{\mathrm{ring}}/J \sim (t_a/t_b)^4$
increases, and the ring term becomes increasingly important in the Hamiltonian,
$H_b = H_J + H_{\mathrm{ring}}$.

Consider the phase diagram for the boson Hamiltonian $H_b$,
when the bosons are at some generic density, $n_b \equiv \langle b^\dagger_i b_i \rangle$.
For $K_{\mathrm{ring}} << J$ one expects a Bose-condensed superfluid
phase.  In Ref.[\onlinecite{DBL}] variational wavefunction studies found that the
D-wave Bose-Metal (DBM) was stabilized for $K_{\mathrm{ring}}/J \gtrsim 2$ and for boson densities
larger than $n_b \gtrsim 0.4$ (or, under particle-hole symmetry, smaller than $0.6$).  
At other densities in this large $K_{\mathrm{ring}}$ regime, phase separation into a (probable CDW) phase at half-filling and a zero density state was found.

The boson superfluid state corresponds to a fully gapped (conventional)
BCS state of the fermions.  However, the DBM phase of the bosons
would correspond to a truly exotic non-superfluid paired state of fermions, with a full gap
for the single fermion excitations, but with uncondensed ``Cooper pairs".
The Cooper-pairs would form a collective state with 
gapless excitations along surfaces in momentum space - a ``Cooper-pair Bose-Metal" (CPBM)!

Since the fermion density, $n= n_\uparrow + n_\downarrow  = 2 n_b$,
this would suggest that the CPBM phase might be present
in the phase diagram in the regime of intermediate $U/t$, strong anisotropy
with $t_b/t_a$ small and for densities near half-filling, $n \gtrsim 0.8$.
As one can see from Fig. 3(b), this is precisely the regime where
the BCS mean-field treatment finds the stable gapless superfluid (G-SF) state.
The G-SF and CPBM phases are dramatically different.
The G-SF state is a superfluid with gapless fermion excitations, while in the CPBM
phase the fermions are fully gapped but the state is {\it not} a superfluid.
Which of these two states, if either or perhaps both, is present in the true phase diagram
of the original attractive Hubbard model is unknown, and certainly worthy of future investigation.  
Recent studies of the boson ring model on the two-leg ladder have found compelling evidence for a quasi-1D Bose-Metal, a ladder descendant
of the 2D DBM state. \cite{DBL_ladders}

 To summarize, we have proposed a model that allows for realizing exotic paired phases 
 of unpolarized fermion mixtures using spin-dependent optical lattices. The main ingredient is the existence of mismatched Fermi surfaces between both hyperfine states.  
We explored the BCS mean-field phase diagram
for two anisotropic Fermi surfaces rotated by 90 degrees respect to one another, checking for the relative stability of the BCS, FFLO, gapless superfluid and normal metal phases.
Other mean-field states such as a nematic superfluid\cite{nematic}, or multi-modal FFLO states.\cite{crystallography,FFLOladder} could be considered in future work.
We also argued that anisotropic Fermi surfaces plus attractive interactions leads to an effective 
model of Cooper-pairs with a  ring-exchange term, that may allow to realize a paired but non-superfluid  Bose-metal phase.
Whether this Cooper-pair Bose-Metal phase appears for the attractive Hubbard model will be explored in future work.  The Fermi surface geometries presented here are just one example that demonstrate the rich physics that emerges
once one allows for spin dependent Fermi surfaces. Other possibilities can also be explored, such as a combination of circular and elliptical Fermi surfaces, with the only requirement that both Fermi surfaces enclose the same area.

We thank P. Zoller for bringing to our attention the possibility of 
using spin-dependent optical lattices, and J. Porto and I. Bloch for discussing the feasibility of the experiment proposed in this paper.  
We are grateful to D.~Huse, F. Heidrich-Meisner and Olexei Motrunich  for fruitful discussions. 
We acknowledge funding from Microsoft Station Q
and the National Science Foundation through the grant DMR-0529399 (MPAF). 




\begin{thebibliography}{99}

\bibitem{Z-K-review} For a review, see: W. Ketterle, M. W. Zwierlein, `` Making, probing and understanding ultracold Fermi gases''. Ultracold Fermi Gases, Proceedings of the International School of Physics ``Enrico Fermi'', Course CLXIV, Varenna, 20 - 30 June 2006, edited by M. Inguscio, W. Ketterle, and C. Salomon (IOS Press, Amsterdam) 2008.

\bibitem{BCS} J. Bardeen, L. N. Cooper, and J. R. Schrieffer, Phys. Rev. {\bf 108}, 1175 (1957).

\bibitem{FF} P. Fulde and A. Ferrell, Phys. Rev. {\bf 135}, A550 (1964).

\bibitem{LO} A. Larkin and Y.N. Ovchinnikov, Zh. Eksp. Teor. Fiz. {\bf 47}, 1136 (1964) [Sov. Phys. JETP {\bf 20}, 762 (1965)].

\bibitem{Radovan Nature 03} H. A.~Radovan, A.~Bianchi, R.~Movshovich, C.~Capan, P. G.~Pagliuso, and J. L.~Sarrao, Nature {\bf 425}, 51 (2003). See also A.~Bianchi, R.~Movshovich, C.~Capan, P. G.~Pagliuso, and J. L.~Sarrao \prl {\bf 91}, 187004 (2003).

\bibitem{experiments1} M. W. Zwierlein {\it et al.}, Science {\bf 311}, 492 (2006); M. W. Zwierlein {\it et al.}, Nature {\bf 442}, 54 (2006); Y. Shin {\it et al.}, \prl {\bf 97}, 030401(2006).

\bibitem{experiments2} G. B. Partridge {\it et al.}, Science {\bf 311}, 503 (2006); G. B. Partridge {\it et al.}, \prl {\bf 97}, 190407 (2006).

\bibitem{FFLOreviews} For a review, see: D. E. Sheehy and L. Radzihovsky, Annals of Physics {\bf 322}, 1790 (2007). See also R. Casalbuoni, and G. Nardulli, \rmp {\bf 76}, 263 (2004).

\bibitem{sarma} G. Sarma, J. Phys. Chem. Solids {\bf 24}, 1029 (1963).

\bibitem{breached1} W. Vincent Liu, and F. Wilczek, \prl {\bf 90}, 047002 (2003).

\bibitem{breached2} M. M. Forbes {\it et al.}, \prl {\bf 94}, 017001 (2005).

\bibitem{rupak} P. F. Bedaque {\it et al}, \prl {\bf 91}, 247002 (2003).

\bibitem{caldas} H. Caldas, \pra {\bf 69}, 0636002 (2004).

\bibitem{Berg08} E. Berg, E. Fradkin, and S. A. Kivelson, \prb {\bf 79}, 064515 (2009).

\bibitem{Berg09} E. Berg, {\it et al.}, arXiv:0901.4826.

\bibitem{mass imbalance} C.A. Stan {\it et al.}, \prl {\bf 93}, 143001 (2004); F. Ferlaino {\it et al.}, \pra {\bf 73}, 040702(R) (2006); C. Ospelkaus {\it et al.}, \prl {\bf 97}, 120402 (2006).


\bibitem{spin-dependent} O. Mandel {\it et al.}, Nature {\bf 425}, 937 (2003). O. Mandel {\it et al.}, \prl {\bf 91}, 010407 (2003).

\bibitem{liu04} W. Vincent Liu {\it at al.}, \pra {\bf 70}, 033603 (2004).

\bibitem{raghu0902.1336} S. Raghu {\it at al.}, arXiv:0902.1336.

\bibitem{lee0902.1337} W.-C. Lee and C. Wu, arXiv:0902.1337.

\bibitem{oganesyan01} V. Oganesyan {\it et al.}, \prb {\bf 64}, 195109 (2001).

\bibitem{Wu07} S. Wu {\it et al.}, Phys. Rev. B {\bf 75}, 115103 (2007).

\bibitem{koponen06} T. Koponen {\it et al.}, New. J. of Phys. {\bf 8} (2006), 179.

\bibitem{DBL} O.~I.~Motrunich, and M.~P.~A.~Fisher, \prb {\bf 75}, 235116 (2007).

\bibitem{DBL_ladders} D. N. Sheng {\it et al.}, \prb {\bf 78}, 054520 (2008).

\bibitem{nematic} A. Sedrakian, {\it et al}, \pra {\bf 72}, 013613 (2005).

\bibitem{crystallography} J. A. Bowers and K. Rajagopal, \prd {\bf 66}, 065002 (2002).

\bibitem{FFLOladder} A.~.E.~Feiguin and F.~Heidrich-Meisner, Phys. Rev. Lett. {\bf 102}, 076403 (2009).

\end{thebibliography}
\end{document}